\documentclass[12pt,a4paper,final]{iopart}

\usepackage{iopams}
\usepackage{graphicx}
\usepackage[breaklinks=true,colorlinks=true,linkcolor=blue,urlcolor=blue,citecolor=blue]{hyperref}

\usepackage{graphicx}
\usepackage{accents}

\def\mbold#1{\mbox{\boldmath $#1$}}

\makeatletter
\def\widebar{\accentset{{\cc@style\underline{\mskip10mu}}}}
\makeatother

\begin{document}

\title[Tetrahedral symmetry in Zirconium nuclei]{Tetrahedral symmetry in Zr nuclei: Calculations of~low-energy excitations with Gogny interaction}

\author{Shingo Tagami and Yoshifumi R Shimizu$^{1}$}
\address{$^1$Department of Physics, Faculty of Sciences, Kyushu University, Fukuoka 812-8581, Japan}


\author{Jerzy Dudek}
\address{Institut Pluridisciplinaire Hubert Curien (IPHC),
IN$_2$P$_3$-CNRS {\em and}~Universit\'e~de~Strasbourg, F-67037 Strasbourg, France }
\ead{Jerzy.Dudek@IPHC.CNRS.Fr}


\begin{abstract}

We report on the results of the calculations of the low energy excitation patterns for three Zirconium isotopes, {\em viz.}~$^{80}$Zr$_{40}$,
$^{96}$Zr$_{56}$ and $^{110}$Zr$_{70}$, reported by other authors to be doubly-magic tetrahedral nuclei (with tetrahedral magic numbers $Z$=40 and $N$=40, 56 and 70). We employ the realistic Gogny effective interactions using three variants of their parametrisation and the particle-number, parity and the angular-momentum projection techniques. We confirm quantitatively that the resulting spectra directly follow the pattern expected from the group theory considerations for the tetrahedral symmetric quantum objects.
We also find out that, for all the nuclei studied, the correlation energy obtained after the angular momentum projection is very large for the tetrahedral deformation as well as other octupole deformations. The lowering of the energies of the resulting configurations is considerable, i.e.~by about 10 MeV or even more, once again confirming the significance of the angular-momentum projections techniques in the mean-field nuclear structure calculations.

\end{abstract}

\vspace{2pc}
\submitto{\JPG}


\section{Introduction}
\label{sec:intro}

Atomic nuclei are bound by the strong, short-range nucleon-nucleon interactions which impose the average nucleon-nucleon center-to-center distances comparable to the nucleon sizes. As a consequence, nuclear spatial density distributions, say $\rho(x,y,z)$, can be considered well localised in space, described/delimited by abstract auxiliary surfaces, here and below denoted with the symbol $\Sigma$. These surfaces can be defined, for the sake of the following discussion, e.g.~by the relation $\rho(x,y,z)=\frac{1}{2}\rho_c$, where $\rho_c$ can be thought of as the density in the nuclear centre. It then follows, as one of the consequences of the short-range of the nuclear interactions, that the nucleonic density falls rapidly to zero, and, by the same token, that the nuclear mean-field potentials acting on any single nucleon, rapidly vanish when the nucleons leave the nuclear interior (i.e.~outside of $\Sigma$).

These arguments allow one to introduce auxiliary geometrical surfaces to facilitate discussing certain {\em global} nuclear properties -- here: the geometrical forms of the nuclear matter distributions which, at this level of the discussion, can be thought of as a simple classical visualisation of the quantum probability density distributions. However, the problem acquires a much better defined quantum mechanical sense when the symmetry aspects can be discussed by combining the symmetry properties contained within a certain quantum formalism (usually represented by a quantum Hamiltonian approximating the properties of the physical system in question) and the theory of the group representations.


\subsection{Remarks about the symmetry aspects within a many-body theory formulation}
\label{Sect.01.01}

In this article we will present the calculations which can be seen as an example of the applications of the nuclear microscopic many-body theories. Our Hamiltonian of choice will be a microscopic nuclear two-body Hamiltonian with the interactions in the Gogny form and the problem will be solved within the mean-field theory formalism. Generally, the better the mean-field theory approximation, the more precise become the theoretical predictions of the experimental data and the more relevant the examining of the geometrical symmetries associated with nuclear surfaces, $\Sigma$, the latter describing not only the symmetries of the spatial distribution of the nuclear {\em densities} but, more importantly, the symmetries of the {\em spatial behaviour of the mean-field potential}, say $\hat{V}_{mf}$, the latter having a direct impact on the single particle levels. In particular, the single particle energy spectra respect the degeneracies which are {\em in principle} measurable\footnote{This way of regarding the experimental chances of discovering the point-group symmetries in nuclei may, unfortunately, be considered illusory because of the very strong polarisation effects in nuclei. Indeed, suppose an originally spherically-symmetric nucleus is polarised by adding a single nucleon which, using the language of the nuclear mean-field theory, slightly deforms the core by introducing the quadrupole (oblate shape) deformation, say
$
 \alpha_{20}\approx -0.02.
$
Straightforward mean-field calculations show that originally $(2j+1)$-degenerate orbitals loose their degeneracy very quickly. As an example, in the case of the cited minute deformation, the energy difference between the extreme $K$-members of the $i_{13/2}$ orbital,
$
 \delta e \equiv e_{13/2,1/2} - e_{13/2,13/2},
$
approaches 2 MeV or more, an estimate which clarifies the next-to-impossibility of discovering the presence of nuclear symmetries through the single-nucleon degeneracy criteria.} and equal to the dimensions of the irreducible representations of $G$. In the case of the tetrahedral symmetry these degeneracies are equal either 4 or 2 independently of the size of the related mass multiple moment $Q_{32}$.

To be able to briefly discuss the symmetry-related physics-problems when aiming at the discovery of the point-group symmetries of the compact nuclear systems, let us consider a point-group $G$ with the elements
$
 \{{g}_1,{g}_2,\,\ldots\,{g}_f\},
$
under action of which the considered surface $\Sigma$ happens to be invariant. The invariance mentioned implies the commutation-relations in the form
$
 [\hat{H}_{mf},\hat{D}({g}_k)]=0,\;\forall\,k,
$
where the operators $\hat{D}({g_k})$ form a representation of the group $G$. When this happens, the group $G$ turns into a symmetry group of the Hamiltonian -- and, as a consequence -- within the mean-field approximation, also into the symmetry group of the considered nuclear system. This offers the possibility of predicting some measurable consequences since the implied single particle wave functions transform as irreducible representations of the group in question and thus the electromagnetic selection rules can be {\em in principle} derived. Such an approach is, however, not straightforward at all because of the mathematical complications which involve constructions of the determinant-type many-body wave functions and the projection on the irreducible representations of the studied symmetry groups in the appropriately constructed many-body spaces.

In the next section we summarise briefly an alternative quantum approach which does not involve the just mentioned mathematical complications. Our strategy followed in the present article will be to profit from the simplicity of these alternative considerations and approach the physical manifestations of the symmetry issue on the semi-quantitative level.

Whereas the above arguments suggest a research way straightforward to follow, in practice the corresponding analysis is complicated by the necessity of analysing experimentally the branching ratios of very specific transitions whose big number is usually not known at the present time\footnote{Suppose, as our calculations suggest, that some nuclei are tetrahedral-symmetric in their ground-states. An exact-tetrahedral symmetry nucleus is characterised by vanishing dipole and quadrupole moments and, as a consequence, the collective dipole- or quadrupole-transitions which {\em usually dominate in the de-excitation spectra} are expected to be simply absent. What is, however, even more important in the present context is that because of the same reasons, many among the `usually populated levels' will simply not be accessible through the heavy-ion population mechanisms which we are accustomed to consider as `usual'. We arrive at an apparent paradox: The tetrahedral symmetry nuclei are evidently non-spherical, their orientation in space can be defined and therefore they are expected to produce the $E_I\sim I(I+1)$ rotational patterns... except that in contrast to all other nuclei, the corresponding levels are {\em not} connected by the usually strong $E2$-transitions -- and this because the underlying quadruple moments vanish as well as the implied collective reduced transition probabilities.}. Therefore demonstrating the presence of non-trivial geometrical symmetries in nuclei requires proposing certain specifically designed experiments most likely on the step-by-step basis and thus  the study of the discussed symmetry problem becomes a lengthy, multi-step process. It is then of importance to calculate beforehand the leading features of the nuclear spectra in order to focus the possible experimental investigation on the verification of the predicted structures such as the characteristic sequences of the low-lying energy levels which can be considered characteristic for the examined symmetry (see below) and verify the  reduced transition probabilities and branching ratios resulting from the symmetry considerations.


\subsection{Remarks about the symmetry aspects within a quantum-rotor theory
            formulation}
\label{Sect.01.02}

Analogues of the collectively rotating tetrahedral-symmetric nuclei are known to exist in nature, in the domain of molecular physics, where the tetrahedral symmetry molecules have been studied for a long time~\cite{HerzA}. The corresponding formalism related to the group-theory aspects uses the fact that the wave-functions, solutions of the tetrahedral-symmetric quantum rotor Hamiltonian of the system composed of an even number of Fermions, must belong to one of the five irreducible representations of the $T_d\,$-group.

Since these aspects are relatively seldom discussed in the nuclear physics literature let us introduce a few underlying notions which will allow appreciating the simplicity of the symmetry discussions when addressing the rotational properties; actually we will limit ourselves to using the conclusions of such a formulation known from the other authors (see below).

The so-called `usual' case of the nuclear quantum rotor is obtained through quantising the classical tri-axial top energy expression in which case the classical angular momenta, $\ell_x$, $\ell_y$ and $\ell_z$ of a rotating uniform tri-axial ellipsoid are replaced by the corresponding operators $\hat{I}_x$,
$\hat{I}_y$ and $\hat{I}_z$, respectively:
\begin{equation}
       \hat{H}
       =
       \frac{{\hat{I}_x}^{2}}{2\,\mathcal{J}_x}
       +
       \frac{{\hat{I}_y}^{2}}{2\,\mathcal{J}_y}
       +
       \frac{{\hat{I}_z}^{2}}{2\,\mathcal{J}_z},
                                                                 \label{eqn.01}
\end{equation}
where the adjustable constants $\mathcal{J}_x$, $\mathcal{J}_y$ and
$\mathcal{J}_z$ are sometimes called `effective moments of inertia'. To be able to construct a quantum rotor Hamiltonian generalised to higher orders in terms of the angular momentum operators it is convenient to introduce spherical-tensor operator-basis in the form
\begin{equation}
      \hat{T}^\lambda_\mu(n)
      =
      \Big(\underbrace{
            \bigg(
                \big(
                     \hat{I} \otimes \hat{I}
                \big)^{2}
                \otimes \ldots \otimes \hat{I}
            \bigg)^{\lambda-1} \otimes
      \hat{I}}_{n~{\rm times}}
      \Big)^\lambda_\mu
                                                                 \label{eqn.02}
\end{equation}
with the help of which any operator constructed out of powers of the angular momentum can be obtained. The notation above employing the symbol ``$\otimes$'' refers to the usual Clebsch-Gordan coupling and
$
 \hat{I}=\{\hat{I}_{-1},\hat{I}_0,\hat{I}_{+1} \}
$
according to standard notation in terms of spherical tensors; each of the corresponding tensors appears as an $n^{\rm th}$-order polynomial in terms of $\hat{I}$. For instance for the so-called generalised quantum rotor Hamiltonian (cf.~ref.~\cite{Misk04}) we find, using somewhat schematic notation, the following general form
\begin{equation}
       \hat{H}
       =
       \sum_{n=0}^\infty \sum_{\lambda}
           \Big\{
                  h_{\lambda 0}^{} \hat{T}^{\lambda}_0(n)
                  +
                  \sum_{\mu=1}^\lambda
                      \left[
                            h_{\lambda\mu}^{} \hat{T}^\lambda_\mu(n)
                            +
                            (-1)^\mu h^*_{\lambda\mu}
                                           \hat{T}^\lambda_{-\mu}(n)
                      \right]
           \Big\},
                                                                 \label{eqn.03}
\end{equation}
where the adjustable constants $h_{\lambda \mu}$ can be constructed in such a way that the operator on the left hand side above is Hermitian. It is then straightforward to verify with the help of the usual Clebsch-Gordan coupling techniques that the lowest-order tetrahedral-symmetry quantum-rotor Hamiltonian takes the form
\begin{equation}
      \hat{H}_{\rm rotor}
      =
      \underbrace{h_{00}^{} \, T^0_0(2)}_{\displaystyle\hat{H}_{\rm sph.}(2)}
      +
      \underbrace{h_{32}^{} \,
      \big[
          T^3_2(3) - T^3_{-2}(3)
          \big]}_{\displaystyle\hat{H}_{{\rm T}_d}(3)}
      =
      \hat{H}_{\rm sph.}(2) + \hat{H}_{{\rm T}_d}(3),
                                                                 \label{eqn.04}
\end{equation}
where $\hat{H}_{\rm sph.}(2)$ and $\hat{H}_{{\rm T}_d}(3)$ are quadratic and third order operators constructed out of components of $\hat{I}$, of the second and third ranks, respectively. The above object is invariant under the symmetry operations of the $T_d$ group since the first term above is a scalar under the rotation group (and thus invariant under any point-group operations) whereas the second term is invariant specifically under the symmetry operations of the $T_d$-group. The parameters of such an Hamiltonian can be adjusted to simulate the nuclear rotational properties, cf.~ref.~\cite{Misk04} for certain mathematical specificities of the problem.

The matrix representations of the Hamiltonian in (\ref{eqn.04}) can be straightforwardly obtained within the basis of the Wigner functions $\{\mathcal{D}^I_{MK}\}$ and the corresponding solutions constructed numerically together with their classification in terms of the five irreducible representations, the scalar ones, $A_1$ and $A_2$, as well as $E$ and $T_1$ and $T_2$. However even without performing such calculations explicitly one may expect that for the doubly-magic tetrahedral-symmetry even-even nuclei the states belonging to the totally-symmetric representation $A_1$ lie low in the energy scale and it becomes of a primary interest to identify the spins and parities of such states on the basis of the group-theoretical approach -- on the one-hand side -- and to calculate the energies of such states microscopically using our mean-field approach, on the other. For the latter - the rest of this article will be devoted to a description of the corresponding results whereas the solution of the former problem is well known (cf.~ref.~\cite{Tagtetra13}, in particular the discussion summarising the relevant considerations in the Appendix A thereof). The $A_1$-representation contains in its low-spin part the sequence of states with the following spin-parity combinations:
\begin{equation}
      I^\pi \to 0^+,3^-,4^+,6^+,6^-,7^-,8^+,9^+,9^-,10^+,10^-,11^-,
                2 \times 12^+, 12^-,\,\ldots\;.
                                                                 \label{eqn.05}
\end{equation}
As it is discussed in more detail below, in all the three doubly-magic tetrahedral-symmetry nuclei, those and only those states are predicted to form the yrast line (alternatively, to lie close to the yrast states) whereas all other states appear higher in the energy scale.

This property underlies once again -- somewhat paradoxically, since at the level of the two-body formulation of the nuclear Hamiltonian there is formally no place for the point-group symmetry concepts -- the importance of the concepts of the geometrical symmetries. The latter are the fundamental tool in the studies of molecular systems and yet their importance can be seen to extend to the realm of the microscopic {\em nuclear} many-body approaches based on the strong interactions. Our results in the form of the energy-vs.-spin diagrams below, in comparison to the sequence in (\ref{eqn.05}), provide an illustration of the above observation.

At the same time these results suggest that investing in the experimental studies of the nuclei in question is very timely, especially from the point of view of the examining of their octupole properties: The reduced $B(E3)$ transitions and, if the absolute values turn out to be too difficult to obtain, in terms of the corresponding relative transition probabilities and branching ratios.


\subsection{Remarks about the framework of this article}
\label{Sect.01.03}

The present article focuses on this type of the calculations which may facilitate discovering of one of those exotic symmetries -- the tetrahedral one, see e.g.~ref.\,\cite{DGM10} and references therein.
With the help of group theory, one can deduce what kind of spin-parity combinations of the nuclear energy levels should be expected e.g.~at and near the yrast line, the mathematical arguments used in the nuclear context fully analogous to the ones known from molecular physics. However, in order to obtain the possibly reliable knowledge of the characteristic features of the spectra and of the transition probabilities, the method used should preferably be based on a realistic microscopic approach, possibly already tested in the past in a different context.

We have recently analysed theoretically the quantum spectra of the tetrahedrally deformed nuclei with doubly-closed tetrahedral-shell configurations~\cite{Tagtetra13}, where both the spherical and inversion symmetries are broken and the efficient quantum number projection method from the most general mean-field states, ref.\,\cite{TS12}, becomes important. It has been confirmed that the calculated low-lying states compose a collective
band built out of states with some specific spin-parity combinations, which is expected by the group theory considerations.
Moreover, when the tetrahedral deformation increases, the character of the yrast sequence changes from the approximately linear energy-vs.-spin dependence to the parabolic one, just like in the case of the increasing quadrupole deformation.

In our previous study~\cite{Tagtetra13}, a schematic separable-type interaction has been used, employing among others a realistic Woods-Saxon mean-field potential. In this article we present the results obtained using Hartree-Fock-Bogoliubov (HFB) method with the Gogny interactions~\cite{DechGog80}.
We focus the discussion on the realistic calculations with the tetrahedral-symmetry solutions. Some preliminary results were published in ref.\,\cite{TSFSD14}.

We believe that our microscopic results discussed below can be seen as realistic many-body theory predictions of the structure of the low-lying spectra in the three nuclei illustrated here which coincide structurally with the spectra predicted for the tetrahedral-symmetry rotors. We also believe that it is very much worthwhile to attempt the experimental identification of the presence of such states in the nuclei which can be reached under the present day experimental conditions - taking into account, as indicated earlier, that the collective $E1$ and $E2$ transitions are most likely non-existing (negligibly small) under the presence of the exact tetrahedral symmetry and thus the only collective transition will be of the octupole nature, as also discussed in some more detail below.



\section{ Method of calculations }
\label{Sect.02}

We have developed a specifically designed computer program to perform the HFB and projection calculations with the Gogny effective interaction. In the present article we follow the techniques of calculations described already in the previous publications, cf.~e.g.~refs.\,\cite{GirGra83,ANgEgid01}, see also \cite{Tagtetra13,TS12} and references therein. In particular, the HFB equations are solved using the expansion in terms of the harmonic oscillator basis.

One of the characteristic features of the Gogny interaction is that the central two-body potential is represented by the sum of terms composed of Gaussians, for which the matrix elements can be calculated analytically using the hypergeometric functions~\cite{EgidRobChas97}. The Coulomb interaction is accurately approximated by the eight Gaussian terms and treated in the same way as the central interaction. The matrix elements of the zero-range term are calculated with the help of the 61-points Gauss-Hermite quadrature, after having verified that the desired accuracy has been attained.

In order to examine the deformation dependence of the nuclear energy, we have used quadratic constraints~\cite{RS80} in terms of the hermitian-symmetrised multipole moment operators,
\begin{equation}
      \hat{Q}_{\lambda\mu}
      \equiv
      \sum_{i=1}^A
           \left\{\textstyle\frac{1}{2}
                 r^\lambda [Y_{\lambda\mu} + (-)^\mu Y_{\lambda-\mu}]
           \right\}_i,
      \quad
      \mu \ge 0,
                                                                 \label{Qmom}
\end{equation}
where $Y_{\lambda\mu}$ denotes the spherical harmonics. If some specific values of constraints are necessary we use the so-called augmented Lagrangian method~\cite{StBNa10}. The $\lambda=1$ center-of-mass constraints
\begin{equation}
      \Bigl\langle \sum_{i=1}^A x_i \Bigr\rangle
      =
      \Bigl\langle \sum_{i=1}^A y_i \Bigr\rangle
      =
      \Bigl\langle \sum_{i=1}^A z_i \Bigr\rangle
      =0
                                                                 \label{eqn.06}
\end{equation}
are always imposed.

We use the deformation parameters defined through
\begin{equation}
      \alpha_{\lambda\mu}
      \equiv
      \frac{4\pi\,\langle \hat{Q}_{\lambda\mu} \rangle}
           {3A\,\widebar{R}^\lambda},
      \quad\mbox{where}\quad
      \widebar{R}
      \equiv
      \sqrt{\frac{5}{3A}
            \Bigl\langle \sum_{i=1}^{A}r_i^2
            \Bigr\rangle}.
                                                                 \label{eqn.07}
\end{equation}
So defined deformation parameters correspond, to the lowest order, to the usual parameterization of the nuclear surfaces represented by the spherical harmonic expansion
\begin{equation}
       R(\theta\varphi)
       =
       R_0\, c_{\rm V}(\{\alpha_{\lambda\mu}\})
       \Bigl[
            1
            +
            \sum_{\lambda\mu}\alpha^*_{\lambda\mu}Y_{\lambda\mu}(\theta\varphi)
       \Bigr],
                                                                 \label{eqn.08}
\end{equation}
where $R_0$ is the nuclear radius and the numerical factor $c_{\rm V}$, a function of the actual set of the deformation parameters
$\{\alpha_{\lambda\mu}\}$, is introduced to guarantee that the nuclear volume does not depend on the deformation. In this article we focus on the results with the axial-quadrupole constraint operators $\hat{Q}_{20}$ (alternatively
$\alpha_{20}$) and the tetrahedral one, the latter given by $\hat{Q}_{32}$ (alternatively $\alpha_{32}=\alpha_{3-2}$), but we also consider the constraints related to other octupole moments $\hat{Q}_{3\mu\neq 2}$.

Traditionally the D1S parameterization~\cite{D1S} of the Gogny interactions has been frequently used in the literature, cf.~ref.\,\cite{D1Srev}. However, a few newer forms of the parameterization, such as D1N~\cite{D1N} and D1M~\cite{D1M}, have been proposed recently. We are going to illustrate the differences between them for the nuclear potential energies within the HFB calculations.

After obtaining the constrained HFB state $|\Phi^{}\rangle$, we perform the full quantum number projection from it to obtain the the projected wave function,
\begin{equation}
     |\Psi_{M;\alpha}^{INZ(\pm)}\rangle
     =
     \sum_{K} g_{K,\alpha}^{INZ(\pm)}\,
              \hat P_{MK}^I \hat P_{\pm}^{}\hat P^N \hat P^Z|\Phi^{}\rangle,
                                                                 \label{eqn.09}
\end{equation}
according to the standard definitions and notation. The amplitude
$g_{K,\alpha}^{INZ(\pm)}$ and the energy eigenvalue $E_\alpha^{INZ(\pm)}$ are obtained by the so-called Hill-Wheeler {\em Ansatz}~\cite{RS80},
\begin{equation}
      \sum_{K'}{\cal H}_{K,K}^{INZ(\pm)}\ g_{K',\alpha}^{INZ(\pm)}
      =
      E_\alpha^{INZ(\pm)}
      \sum_{K'}{\cal N}_{K,K'}^{INZ(\pm)}\ g_{K',\alpha}^{INZ(\pm)},
                                                                 \label{eqn.10}
\end{equation}
where the kernels are defined by
\begin{equation}
   \left\{
   \begin{array}{c}
         {\cal H}_{K,K'}^{INZ(\pm)} \\[2mm]
         {\cal N}_{K,K'}^{INZ(\pm)}
   \end{array}
   \right\}
   =
   \langle \Phi |
   \left\{
   \begin{array}{c}
          \hat{H} \\[2mm]
          1
   \end{array}
   \right\}
   \hat{P}_{KK'}^I \hat{P}^N \hat{P}^Z \hat{P}_{\pm}
   | \Phi \rangle.
                                                                 \label{eqn.11}
\end{equation}

Since the tetrahedral deformation breaks both the axial and inversion symmetries, we will perform the final calculations using the three dimensional angular momentum and parity projections, $\hat P_{MK}^I$ and $\hat P_{\pm}^{}$, simultaneously - unless - for the comparison purposes, the angular-momentum projection will be switched off. The particle number projection (neutrons,
$\hat P^N$, and/or protons, $\hat P^Z$) is optionally employed if the system is in the superfluid phase. There are, however, ambiguities for the treatment of the density-dependent term to evaluate the Hamiltonian kernel, see e.g.~ref.\,\cite{DDRob10}. We follow Ref.~\cite{RodEgid10} and adopt the projected density prescription for the number projection~\cite{ANgEgid01a} and the conventional mixed (or transition) density prescription for the angular momentum and parity projections.

Concerning some specificities of our projection calculations, for more details c.f.~\cite{TS12}, the value of the basis cut-off parameter is chosen at the level of $10^{-6}$, i.e., only the canonical basis states whose occupation probabilities are larger than $10^{-6}$ are considered. For solving the Hill-Wheeler equation, the norm cut-off parameter $10^{-8}$ is used, i.e., the eigenstates of the norm-kernel whose norm eigenvalues are smaller than $10^{-8}$ are excluded. As it has been found out in ref.\,\cite{TS12}, it is important to include the time-odd components in the HFB state in order to obtain a reliable estimate of the moment of inertia.
For this purpose we employ a small perturbation term, ref.\,\cite{TS12,Tagtetra13}, in the form which resembles the cranking Hamiltonian,
\begin{equation}
      H'=H-\omega_{\rm rot}\,\mbold{n}\cdot\mbold{J},
                                                                 \label{eqn.12}
\end{equation}
where we use $\hbar\omega_{\rm rot}=20$ keV and the direction vector $\mbold{n}$ along the $y$-axis in the present work. It should be noticed that the results do not depend on the actual value of the frequency parameter $\omega_{\rm rot}$, as long as its value is small enough~\cite{TS12}. Note, however, that one cannot use too small a frequency, because then the time-odd components are too small and excluded by the norm cut-off: The value 20 keV is a result of a compromise between these two factors. It has also been confirmed in
ref.~\cite{Tagtetra13} that for the non-zero tetrahedral deformation the calculated result does not depend on the choice of direction vector $\mbold{n}$.


\section{ Results of the calculations }
\label{Sect.03}

A spherically-symmetric Cartesian harmonic oscillator basis is employed with the oscillator frequencies $\omega_x=\omega_y=\omega_z$, the condition required to perform the accurate angular momentum projection.  All the basis states with the oscillator quantum numbers $(n_x,n_y,n_z)$, with
$
 n_x+n_y+n_z \le N_{\rm osc}^{\rm max}
$
are retained.

\begin{figure}[!ht]
\begin{center}
\includegraphics[width=80mm]{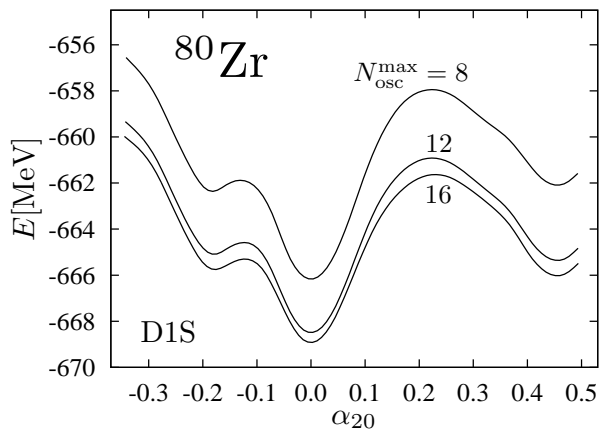}
\caption{
         Convergence properties in the constrained HFB calculations
         (no projection) with the
         Gogny D1S parameterisation for $^{80}$Zr as a function of the
         deformation parameter $\alpha_{20}$ for an increasing model space
         defined by the maximum oscillator shells $N_{\rm osc}^{\rm max}=$8,
         12, and 16.
}
                                                                 \label{fig.01}
\end{center}
\end{figure}

An example of the results of the Gogny-HFB calculations with the axial-quadrupole constraint ($r^2Y_{20}$) is shown in fig.\,\ref{fig.01} in the form of functions of the deformation parameter $\alpha_{20}$ for $^{80}$Zr nucleus using the D1S parametrisation of the Gogny interaction Hamiltonian. The results using the model space specified by $N_{\rm osc}^{\rm max}=8$, 12, and 16 are compared. It can be seen that the energy differences between the spherical and
prolate minima are larger by about 1 MeV in the calculation with
$
 N_{\rm osc}^{\rm max}=8
$
than in that with $N_{\rm osc}^{\rm max}=16$. It is therefore desirable to use as large $N_{\rm osc}^{\rm max}$ as possible, but the large model space requires large computational effort, especially for the quantum number projection calculations. In this article we present the results primarily with
$
 N_{\rm osc}^{\rm max}=12
$
(i.e.~13 shells) or  $8$ (i.e.~9 shells) as compromise choices, verifying each time that the acceptable stability of the final result with respect to the basis cut off has been achieved.

One of the main purposes of the present article is to study the spectroscopic properties of the tetrahedral `doubly-magic' nuclei which happen to carry the `ideal' tetrahedral deformation. For this purpose we choose the tetrahedral doubly magic nuclei of ref.\,\cite{Dudek02},
$^{80}_{40}$Zr$_{40}^{}$,
$^{96}_{40}$Zr$_{56}^{}$, and
$^{110}_{\;\;40}$Zr$_{70}^{}$. They contain relatively small numbers of nucleons and are therefore relatively less demanding for the calculations with the full quantum number projections.


\subsection{Tetrahedral deformation effects in the $^{80}_{40}$Zr$_{40}$
            nucleus}
\label{Sect.03.01}

\begin{figure}[!ht]
\begin{center}
\includegraphics[width=80mm]{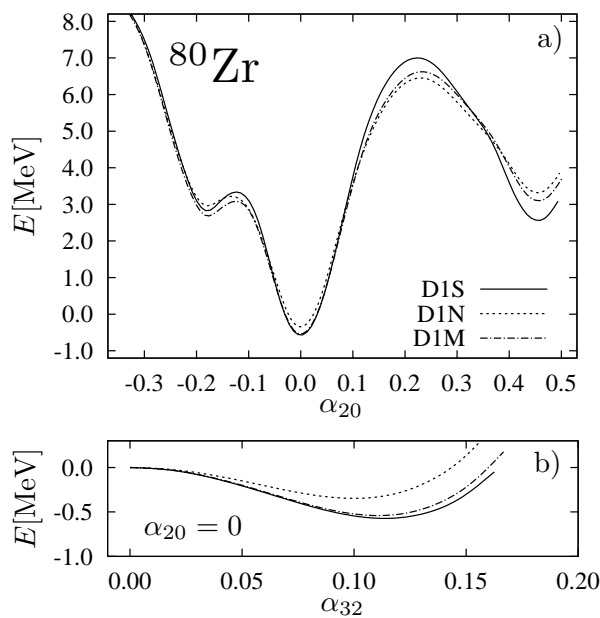}
\caption{
         Potential energy curves resulting from the constrained Gogny-HFB
         calculations as functions of the deformation parameter $\alpha_{20}$
         (top) and of the deformation parameter $\alpha_{32}$ (bottom) for
         $^{80}$Zr ($Z=N=40$). The results with three variants of the
         Gogny interactions labelled, D1S, D1N, and D1M are compared. The model
         space is defined by $N_{\rm osc}^{\rm max}=$12.
}
                                                                 \label{fig.02}
\end{center}
\end{figure}

Potential energy curves for $^{80}$Zr, calculated with the constrained Gogny-HFB  approach are shown in fig.\,\ref{fig.02}. We employ three variants of the parameterization of the Gogny interaction Hamiltonian: D1S, D1N, and D1M. Here and in the following, we choose the origin of the energy scale (zero energy) as the spherical HFB energy (without projection) for each case. As can be seen from the figure, all the three variants of the parameterisation give similar potential energy curves. Our prediction is (cf.~the bottom panel of the figure) that the ground-state has a pure tetrahedral deformation with $\alpha_{20}=0$ and $\alpha_{32} \ne 0$; its energy lies approximately $0.6$~MeV below the energy of spherical-shape configuration.

We found that the single-particle energy gaps with $Z=N=40$ are very large so that the pairing correlations vanish for both the neutrons and the protons. There are a few HFB or HF-BCS calculations with zero-range Skyrme interactions for the tetrahedral deformation in this nucleus~\cite{TYM98,YMM01,ZbMH06}. Compared with the result of HF-BCS with the SIII interaction in ref.~\cite{TYM98}, the energy gain of the tetrahedral minimum in our calculations is slightly smaller, while it is larger than that of refs.\,\cite{YMM01,ZbMH06}. It should be noted that the definition of our deformation parameters is different from those in the cited references; their deformation parameter corresponding to our $\alpha_{\lambda\mu}$ is a factor  $\sqrt{2}$ larger for $\mu \ne 0$.
Moreover the other authors use $\widebar{R}$=1.2$A^{1/3}$ fm instead of the calculated radius parameter as in eq.\,(\ref{eqn.07}). We believe that our definition is more appropriate to calculate the magnitude of the deformation, especially when the specific nuclear structure effects such as skin and/or halo become important for certain nuclei.

We have tested that the obtained tetrahedral-deformed configuration is really the minimum by restarting the HFB iterations from initial states which are perturbed in such a way that they have small extra moments
$
 \langle \hat{Q}_{\lambda\mu}\rangle
$
with $2 \le \lambda \le 4$ and $\mu \le 4$ and break the tetrahedral symmetry.  After performing a sufficiently high number of iterations, the obtained minimum deformations coincide with the original tetrahedral-symmetric solution.

The higher-rank $\lambda$ non-zero multipole moments of the present HFB minimum with the D1S parametrisation are listed in Table~\ref{tab.01}. The lowest order tetrahedral-symmetry multipole-moment is of the rank $\lambda=3$ and is specified by the only non-zero component $Q_{32}$. The next order tetrahedral-symmetry moments allowed by the group considerations must have the rank $\lambda=7$. The tetrahedral-symmetry at this rank is specified by a simultaneous combination of two moments {\em viz.} $Q_{72}$ and $Q_{76}$ related by a fixed coefficient, i.e.: $Q_{76}=-Q_{72}\times\sqrt{{11}/{13}}$ (cf.~ref.~\cite{DDD07}). As it can be seen from the table, this geometrical symmetry relation is verified up to a six digit accuracy illustrating at the same time certain aspects of precision achieved in our numerical calculations.

Recall that the tetrahedral group is a subgroup of the octahedral group, and therefore the tetrahedral-symmetry states are characterised by a simultaneous presence of both the tetrahedral and octahedral deformations. In other words, the non-zero tetrahedral-symmetry moments at the tetrahedral-symmetry minima are usually accompanied by finite octahedral-symmetry moments~\cite{DDD07}. Let us remind the reader at this point that, after the cited reference, the lowest order octahedral deformation has the rank $\lambda=4$ and is determined by the combined $Q_{40}$ and $Q_{4\pm4}$ multipole moments according to the relation
$
 Q_{4\pm 4}=-Q_{40}\times\sqrt{{5}/{14}}.
$
The next order octahedral-symmetry moments are of the rank $\lambda=6$ and are determined by a combination of $Q_{60}$ and $Q_{6\pm4}=Q_{60}\times\sqrt{{7}/{2}}$ which are the only allowed non-zero multipole moments of this rank. As it can be seen from Table~\ref{tab.01}, the tetrahedral-symmetry properties of the minimum configuration are perfectly respected by our HFB calculations represented in terms of the mass multiple moments with the (at least) six decimal digit accuracy. [Similar analysis has been performed for the selfconsistent Skyrme HFB calculations in Ref.~\cite{DDD07}.]

\begin{table}[ht]
\begin{center}
\begin{tabular}{cccccc}
 $Q_{32}$ & $Q_{72}$ & $Q_{76}$ & $-Q_{72}\times \sqrt{\frac{11}{13}}$ & & \cr
\hline
$354.970$ & $-8897.50$ & $8184.51$ & $8184.51$ & &
\vspace*{5mm} \cr
$Q_{40}$ & $Q_{44}$ & $-Q_{40}\times\sqrt{\frac{5}{14}}$ &
$Q_{60}$ & $Q_{64}$ & $Q_{60}\times\sqrt{\frac{7}{2}}$ \cr
\hline
$-283.937$ & $169.685$ & $169.685$ & $5195.58$ & $9720.03$ & $9720.03$ \cr
\end{tabular}
\end{center}
\caption{
         The non-zero multipole moments $Q_{\lambda\mu} $ (with $\lambda \le 7$)
         in unit of [fm$^\lambda$] for the tetrahedral HFB minimum state
         obtained using the D1S parametrisation in $^{80}$Zr.
         The upper (lower) tables refer to the
         tetrahedral (octahedral) degrees of freedom; for more details see the
         text.
}
                                                                 \label{tab.01}
\end{table}
\begin{table}[ht]
\begin{center}
\begin{tabular}{cccccc}
$\alpha_{32}$ & $\alpha_{72}$ & $\alpha_{76}$ &
$-\alpha_{72}\times \sqrt{\frac{11}{13}}$ & & \cr
\hline
$0.1139344$ & $-0.003047097$ & $0.002802922$
& $0.002802922$ & &
\vspace*{5mm} \cr
$\alpha_{40}$ & $\alpha_{44}$ & $-\alpha_{40}\times\sqrt{\frac{5}{14}}$ &
$\alpha_{60}$ & $\alpha_{64}$ & $\alpha_{60}\times\sqrt{\frac{7}{2}}$ \cr
\hline
$-0.01646264$ & $0.009838308$ & $0.009838309$ &
$0.01054496$ & $0.01972782$ & $0.01972781$ \cr
\end{tabular}
\end{center}
\caption{
         Similar to the above but for the non-zero deformation parameters
         $ \alpha_{\lambda\mu} $ ($\lambda \le 7$) for the tetrahedral HFB
         minimum state obtained by the D1S parametrisation in $^{80}$Zr.
         The upper (lower) tables refer to the tetrahedral (octahedral)
         deformation. The nine digit precision is used not so much to represent
         the absolute values of the results but rather the precision with which
         the tetrahedral symmetry is respected by our self-consistently
         calculated wave functions.
}
                                                                 \label{tab.02}
\end{table}
Since the multipole moments of the increasing rank $\lambda$ are usually given by quickly increasing numbers, as seen in table \ref{tab.01}, it is rather difficult to imagine the information about shapes of the nuclear surfaces which such moments carry. Because of this it will be convenient to represent the same information using the nuclear deformations introduced by eqs.\,(\ref{eqn.07}), see also the surrounding text. This is done with the help of table \ref{tab.02} showing that within at least nine decimal places the numerical results do represent the mathematical correlations expected on the basis of the group theory considerations.

Recently, the state-of-the-art quantum number projection and configuration mixing calculations for quadrupole degrees of freedom have been performed in ref.\,\cite{RodEgid11} with the Gogny D1S interaction for the same nucleus.
The reflection symmetry has been assumed and no tetrahedral deformation
has been taken into account in Ref.~\cite{RodEgid11}.
In the latter reference, the mean-field states determined by variation after number-projection calculation (PNP-VAP) are employed to calculate the potential energy curves rather than the HFB states,
but the results of the energy curves are similar to ours: The state with no quadrupole deformation $\alpha_{20}=0$ is the lowest one in this case since no octupole deformations were considered and the prolate and oblate minima appear as excited states.

Let us mention that the form of the curves and in particular the prolate and oblate deformation minima in ref.\,\cite{RodEgid11} may seem different from ours. However, we have verified that this is because of the different definition of the deformation parameter, i.e., $\widebar{R}$=1.2\,$A^{1/3}$ fm used in
ref.\,\cite{RodEgid11} in their definition instead of ours in
eq.\,(\ref{eqn.07}). In particular our calculated r.m.s.~radii for $^{80}$Zr are slightly larger, the obtained values being $\widebar{R}$=1.243\,$A^{1/3}$ fm at the spherical HFB state and $\widebar{R}$=1.321\,$A^{1/3}$ fm at the near super-deformation of $\alpha_{20}$=0.563. Although the spherical minimum is the lowest in energy with the PNP-VAP calculation, the effect of the angular momentum projection is so large that the quadrupole deformed state becomes the ground state according to calculations including the projection and configuration mixing, which seems to be consistent with the experimental data\,\cite{List87}. The correlation energy gained by the angular momentum projection is shown to be down by $(4 - 6)$ MeV for various quadrupole deformed states in ref.\,\cite{RodEgid11}, which is confirmed also in our calculations.
\begin{figure}[!ht]
\begin{center}
\includegraphics[width=80mm]{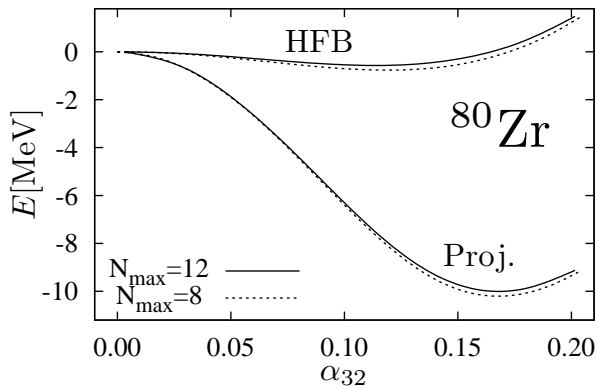}
\caption{
         The projected energy curves for $I^\pi=0^+$ state as
         functions of tetrahedral deformation for $^{80}$Zr - compared with the
         HFB energy curve. The D1S parametrisation
         has been employed in this case; the results of calculations with two
         model spaces defined by $N_{\rm osc}^{\rm max}=$8 and 12 are presented.
}
                                                                \label{fig.03}
\end{center}
\end{figure}

It is very instructive to examine the effects of the projection also for the tetrahedral deformation. The energies of the $I^\pi=0^+$ state have been calculated using the projection technique from the HFB states subject to the
$\hat{Q}_{32}$-constraint as a function of $\alpha_{32}$. The results are shown in fig.\,\ref{fig.03}. In these calculations, the numbers of the mesh points in the Gaussian quadratures of the projectors are
$
 (N_\alpha=N_\gamma,N_\beta)=(42,42)
$
for the Euler angles $(\alpha,\beta,\gamma)$. There are large single-particle energy gaps at the non-zero tetrahedral deformation for both the neutrons and protons at the particle number $N=Z=40$. The tetrahedral-minimum Gogny-HFB states in $^{80}$Zr are thus not superfluid, both for the neutrons and protons, so that the particle-number projection is not necessary in this particular case.

In the discussed case only the results with the D1S interaction are presented  because those with the two other variants of the parameterisation are very similar. As it is seen from fig.\,\ref{fig.03}, where the $I^\pi=0^+$ projected energy curves are plotted as functions of $\alpha_{32}$, the correlation energy gain relative to the spherical configuration -- due to the angular momentum projection -- is very large and amounts to about 10~MeV for the tetrahedral deformation
$\alpha_{32}\approx 0.17$, the effects which are much larger than those for the quadrupole deformations. In fig.\,\ref{fig.03} we have included the two results with $N_{\rm osc}^{\rm max}=$8 and 12; it is seen that the correlation energy of the quantum number projection can be accurately estimated with the smaller model space $N_{\rm osc}^{\rm max}=8$.

\begin{figure}[!ht]
\begin{center}
\includegraphics[width=140mm]{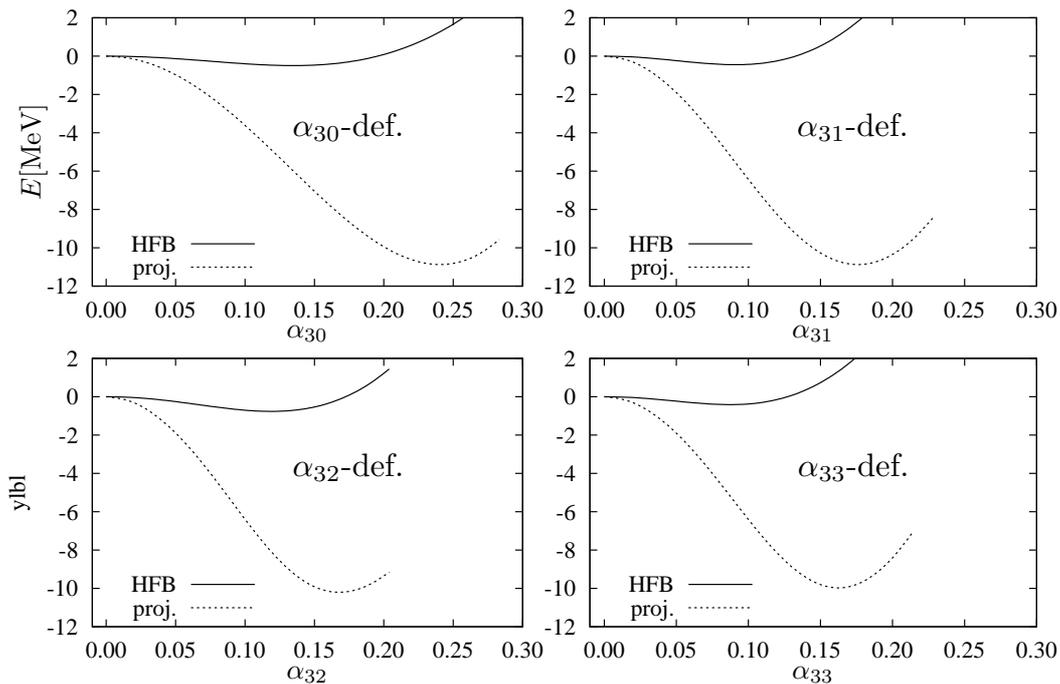}
\caption{
         The HFB and projected energy curves for $^{80}$Zr as in
         fig.\,\ref{fig.03} but applying the single-constraint condition for
         each of octupole components $\alpha_{3\mu}$ ($\mu=0,1,2,3$) as
         discussed in the text. The model space is $N_{\rm osc}^{\rm max}=$8
         in this case.
}
                                                                \label{fig.04}
\end{center}
\end{figure}

The experimentally known octupole reduced transition probabilities in the Zirconium mass-region are about the largest ever measured in nuclear physics with the value for $^{96}$Zr reaching as much as $B(E3)$=57$\pm$4
W.\,u.~(cf.~ref.~\cite{Dudetal14} for a comparative table and the original references). The predicted tetrahedral configurations (non-zero octupole
$\alpha_{32}$ deformations with vanishing quadrupole deformation components) are very low-lying in the energy scale and in some of these nuclei are predicted to be very likely tetrahedral-deformed in their ground-states, \cite{Dudetal14}. This remark has a couple of consequences whose consideration may be instructive. Firstly, if the considered tetrahedral configurations are indeed ground-state ones the generally difficult measurements of the reduced transition probabilities in question needed for further study of the discussed phenomena may turn out to be slightly easier. Secondly, the large $B(E3)$ values come in this case principally from `pure' octupole contributions -- as opposed to the situation in which octupole {\em and} quadrupole deformations coexist and it becomes of particular interest to examine the possible effects of all the
$\hat{Q}_{3\mu\neq 2}$ moments to the final reduced transition probabilities. With this goal in mind it becomes instructive to test the total energy behaviour in function of the other octupole-deformation parameters.

We have calculated the HFB as well as the projected energies in function of these components treated as constraints for $^{80}$Zr. The results are shown
in fig.\,\ref{fig.04}. They are obtained by using the smaller model space with $N_{\rm osc}^{\rm max}=$8 (for the sake of completeness, the result for the tetrahedral deformation $\alpha_{32}$, which is the same as fig.\,\ref{fig.03}, is also included).

In order to be able to extract the sought information in a way which resembles the one-axis projection cuts like those presented so far we have imposed a single active-constraint condition e.g., for calculating the HFB energy as a function of $\alpha_{31}$, the components $\hat{Q}_{3\mu}$ with $\mu \ne 1$ are constraint to be zero and then the angular-momentum projection is applied. For all the four constraints studied in the case of $^{80}$Zr, the HFB energy gain relative to the spherical configuration is similar, but the largest one corresponds to the tetrahedral deformation.  The latter defines a real local HFB minimum at $\alpha_{32}=0.114$ which turns out to be the absolute one, according to our calculations.
Compared with this minimum energy, the lowest energies
for other $\alpha_{30}$, $\alpha_{31}$ and $\alpha_{33}$ constraints
are 272, 318 and 354 keV higher, respectively.
We may associate this feature with the presence of the four-fold degeneracies of the single-nucleon levels at the tetrahedral-symmetric shapes as implied by the presence of the four-dimensional irreducible representations of the tetrahedral group -- what results in stronger (tetrahedral) shell-effects and stronger chances to generate the bigger level spacings.

It may be instructive to discuss a possible impact of the quadrupole
deformation developing along the paths corresponding to
the $\alpha_{3K}$ constraints discussed above.
As for the tetrahedral deformation, $\alpha_{32}$, the quadrupole deformation
is always vanishing thus the tetrahedral symmetry is preserved along
the corresponding static minimum path.
For the $\alpha_{30}$ constraint, only $\alpha_{20}$ is non-vanishing
and gradually increases up to about 0.045 at the largest value of constraint
in fig.\,\ref{fig.04}.  Similarly,
for the $\alpha_{31}$ constraint, both $\alpha_{20}$ and $\alpha_{22}$
gradually increase up to about 0.05 and 0.035, respectively;
for the $\alpha_{33}$ constraint, only $\alpha_{20}$ gradually
decreases down to about $-0.07$.
Thus, one can safely say that the effect of the static quadrupole deformation
is minor.

The analogous results for the angular-momentum projected energies are similar among themselves. The obtained lowering of the energy due to projection is very significant: indeed, the projected nuclear energies are lowered by about 10~MeV {\em for each} of the four deformations as illustrated in fig.\,\ref{fig.04}.
Among the lowest projected energies along the paths corresponding to
the four constraints, those for the $\alpha_{30}$ and $\alpha_{31}$
constraints coincide within 1 keV and the others are 673 and 901 keV higher,
respectively, for the $\alpha_{32}$ and $\alpha_{33}$ constraints.
We conclude that in the nucleus studied {\em all the octupole deformation components are comparably important} according to our projected Gogny-HFB calculations.

The above presented results for the HFB local minima can be compared with those in refs.~\cite{TYM98,YMM01,ZbMH06}, the latter suggesting as well that the nucleus $^{80}$Zr is soft against not only the tetrahedral deformation ($\alpha_{32}$) but also other components of the octupole deformation.


\subsection{Low-spin low-energy levels in the $^{80}_{40}$Zr$_{40}$ nucleus}
\label{Sect.03.02}

To prepare the way for the possible experimental tests of the symmetry properties introduced so far, one of the first issues which comes to one's mind is that of the spin-parity sequence of the energy levels in their consecutive appearance when spin increases, especially at- and close to the yrast line where the excited levels are usually populated most easily. In this context it will be of particular importance to examine such features expected for the pure tetrahedral symmetry as an ideal limiting case when trying to obtain the experimental confirmation of the presence of the discussed symmetry in nuclei. In our previous study~\cite{Tagtetra13}, we have investigated the $E$-vs.-$I$ pattern of the lowest collective excitations as a function of the tetrahedral deformation. It has been shown that at small deformations the energy vs.~spin sequence is approximately linear, whereas at increasing deformation the energy vs.~spin relation approaches the usual parabolic form. Qualitatively, according to our calculations, the deformations at which a transition between the two patterns take place get relatively smaller with increasing nuclear mass.

\begin{figure}[!ht]
\begin{center}
\includegraphics[width=70mm]{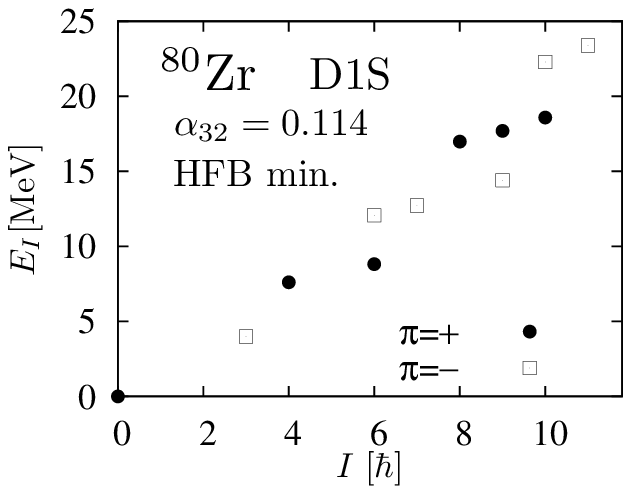}
\includegraphics[width=70mm]{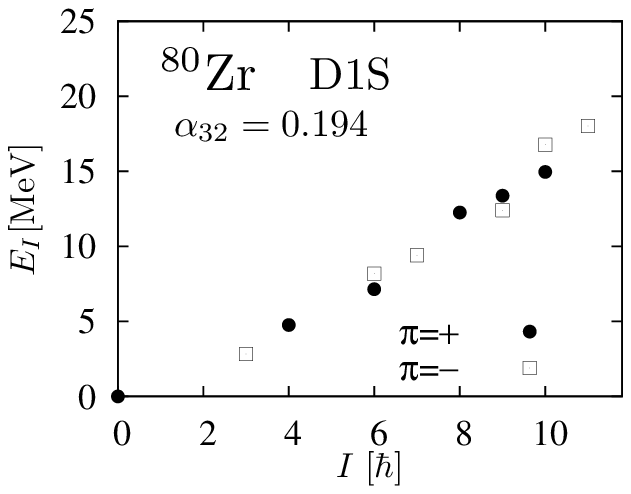}
\caption{
         Lowest-lying energy levels for $^{80}$Zr in function of spin
         calculated by the projection method from the HFB minimum state
         with $\alpha_{32}=0.114$, left, and from a constrained HFB state with
         $\langle \hat{Q}_{32}\rangle=600$ fm$^3$ corresponding to
         a larger deformation, $\alpha_{32}=$0.194, right.
		 The Gogny D1S interaction is used with the model space defined by
         $N_{\rm osc}^{\rm max}=12$.  Here and in the following
         (except fig.\,\ref{fig.06}) the effect of the perturbative time-odd
         term discussed in the text, c.f.~eq.\,(\ref{eqn.10}),
         with $\hbar\omega =20$ keV has been included.
         [For the comments about
         certain characteristic features of these diagrams see the text.]
}
                                                                 \label{fig.05}
\end{center}
\end{figure}

The excitation-energy spectrum obtained by the projection technique applied at the HFB minimum is shown in fig.\,\ref{fig.05}, left. The D1S parameterization has been used here. The tetrahedral deformation at the minimum is relatively small, $\alpha_{32}=$0.114, and thus the spectrum has approximately linear energy vs.~spin dependence. For comparison we have calculated the analogous excitation spectrum for a larger tetrahedral deformation assuming
$
 \langle \hat{Q}_{32}\rangle=600$ fm$^3,
$
which corresponds to $\alpha_{32}=$0.194, cf.~fig.\,\ref{fig.05}, right. At the latter deformation, the resulting spectrum has still a `transitional character'
i.e.~a nearly linear energy vs.~spin dependence -- but the fluctuations of the level positions around an imaginary straight line representing the average level positions are weaker.

A couple of observations need to be emphasised. Firstly, the levels shown in fig.\,\ref{fig.05} represent {\em all} the low-energy states which either form the yrast line or lie close to it. All levels not shown in the figure, in particular the states at spins $I=1,2$ and 5\,$\hbar$, lie much higher in the energy scale forming a sequence roughly parallel to the approximately linear sequences shown; the latter is positioned approximately 6-to-8 MeV higher in the energy scale! This form of the behaviour has been presented and discussed in some detail earlier and the interested reader is referred to \cite{Tagtetra12}. Let us stress that the levels `coming down' in our Gogny-HFB calculations (the ones which are found in the figure) are exactly those characterised by the spin-parity combinations characteristic of the $A_1$ irreducible representation of the tetrahedral group as listed in eq.\,(\ref{eqn.05}). This observation can be seen as the sign of high relevance and usefulness of the notions of point-group representation-theory in that the characteristic features of the spectra can be traced back to the information about the irreducible representations of the groups in question.

Secondly, let us stress that our theoretical predictions are limited by the fact that they are obtained within the mean-field theory and are focussed on the tetrahedral symmetry minima. Thus they do not include any information about neither collective rotation effects associated with the higher-lying super-deformed or oblate-shape configurations nor e.g.~the low-lying 1-particle 1-hole excitations which are expected to form the characteristic sequences with the spins $I=j^2=0,2,4,\,\ldots(2j-1)$; in fact the latter sequences are known experimentally in several nuclei in the Zirconium region. Similarly to those latter ones, other particle-hole excitations, such as $I\leftrightarrow (j_1\otimes j_2)_I$ are expected to lie relatively low in the energy scale and possibly compete with the states to which our calculations are limited.

\begin{figure}[!ht]
\begin{center}
\includegraphics[width=70mm]{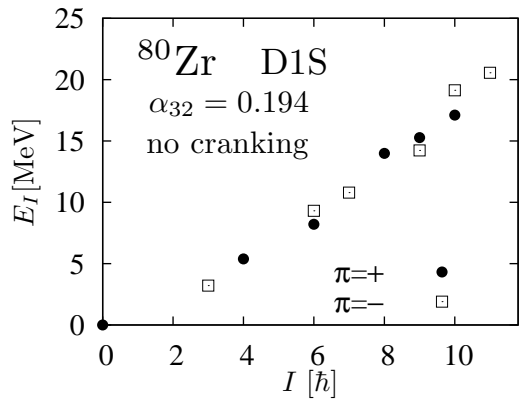}
\caption{
         Similar to the one in fig\,\ref{fig.05}, right, but without
         the perturbative time-odd term (for the definition of this term see
         text). Comparison shows that although the overall structure of the two
         spectra is very similar -- nevertheless the energies of the
         calculations with including the perturbative time-odd term lie
         noticeably lower in energy.
}
                                                                \label{fig.06}
\end{center}
\end{figure}

In order to illustrate the influence of the perturbative time-odd term in the Hamiltonian which resembles the cranking term, the result of projection from the HFB state without the `cranking' are shown in fig.\,\ref{fig.06}. The slope of the excitation energy with respect to spin is increased by roughly about 15\% through this inclusion of the time-odd term.
This effect is not so large compared to the case of the rotational spectra in $^{164}$Er studied in Ref.~\cite{TS12}, where the moment of inertia increases by about 50\% when the effect of `cranking' is taken into account.

In our previous work~\cite{Tagtetra13}, we have employed a rather simple parameterisation of the Hamiltonian in which the single-particle potential has been chosen in the form of the universal Woods-Saxon potential combined with the residual interactions of separable multipole-multipole type.  This latter interaction is composed of terms with multipolarity $\lambda=$2, 3, and 4 for the particle-hole channel and with $\lambda=$0 and 2 for the particle-particle (pairing) channel. The radial form-factors have been chosen of the surface type,
i.e.~proportional to the derivative of the Woods-Saxon potential, for the particle-hole channel, and the volume-type, i.e.~proportional to the usual
$r^\lambda Y_{\lambda\mu}$ factors, for the pairing channel. The strengths of the interactions for the particle-hole channel have been chosen selecting the so-called self-consistent value as in ref.\,\cite{BM75}, while the strength of the monopole pairing interaction has been determined in such a way as to reproduce the observed even-odd mass difference.

\begin{figure}[!ht]
\begin{center}
\includegraphics[width=70mm]{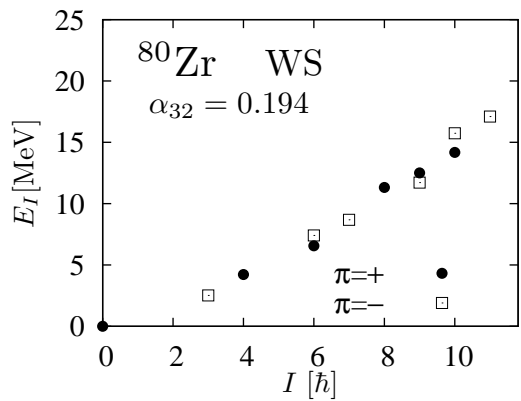}
\caption{
         Results similar to those in fig.\,\ref{fig.05}, right, here calculated
         by employing a simplified model with the Woods-Saxon potential combined
         with the separable multipole-multipole interactions as in
         ref.\,\cite{Tagtetra13} instead of the Gogny interactions.
}
                                                                \label{fig.07}
\end{center}
\end{figure}

The ratio of the monopole and quadrupole interactions has been chosen as 13.5
(see ref.~\cite{Tagtetra13} for details), what reproduces satisfactorily the moments of inertia for nuclei in the rare earth region. Figure~\ref{fig.07} illustrates the spectrum calculated with this briefly recapitulated scheme of ref.\,\cite{Tagtetra13}, using the tetrahedral deformation equal that of the projected minimum with the Gogny D1S interaction. As it is seen the calculated
spectrum is quantitatively similar. Comparison shows that the rotational excitations of the tetrahedral-symmetry nuclei are not much sensitive to the detailed form of the effective interaction and that the schematic interaction can describe them rather well; This has been also known for the collective spectra of quadrupole deformed systems~\cite{BK68}.

\begin{figure}[!ht]
\begin{center}
\includegraphics[width=70mm]{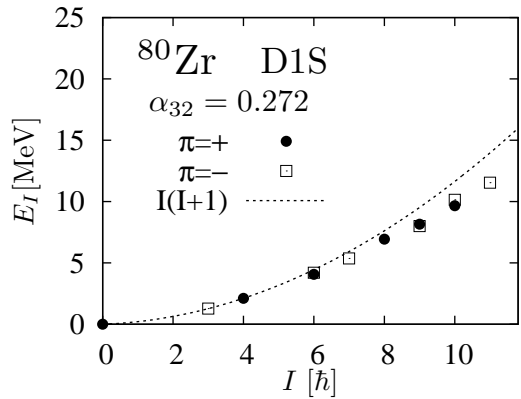}
\caption{
         The calculated lowest energy sequences for both parities, employing the
         projection and the Gogny D1S interaction, but for a larger deformation
         $\alpha_{32}=0.272$ corresponding to the constraint value
         $ \langle\hat{Q}_{32}\rangle=900$ fm$^3$.
		 The dotted line represents a~reference $I(I+1)$ parabola passing
		 through the origin and the first excited $3^-$ state.
}
                                                                 \label{fig.08}
\end{center}
\end{figure}

It is instructive to examine the evolution of the rotational spectra at increasing tetrahedral deformation with the Gogny interactions as compared with the simplified interactions of ref.~\cite{Tagtetra13}. In fig.\,\ref{fig.08} we show an example of the calculated spectrum using the constraint value
$
 \langle \hat{Q}_{32}\rangle=900$ fm$^3,
$
corresponding to $\alpha_{32}=0.272$, the deformation which is considerably larger than the deformation associated with the calculated HFB minimum.
As it can be seen from the figure, the first $3^-$ energy is considerably lowered at this large deformation, and the states with the same spin with opposite parities are almost degenerate composing a sequence very close to the `usual' $\sim I(I+1)$-parabola. We have verified that the results are very similar to the ones obtained with the simple model of ref.~\cite{Tagtetra13} as long as the same deformations are used.


\subsection{The case of the $^{96}_{40}$Zr$_{56}$ nucleus}
\label{Sect.03.03}

\begin{figure}[!ht]
\begin{center}
\includegraphics[width=80mm]{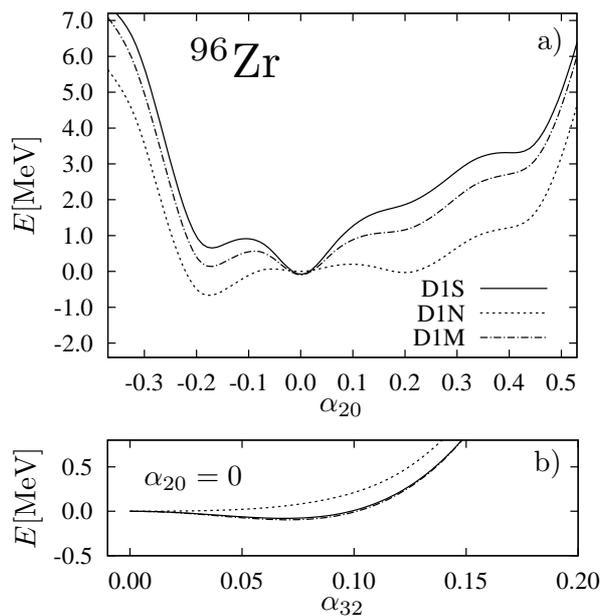}
\caption{
         Potential energy curves resulting from the constrained HFB calculations
         as functions of the deformation $\alpha_{20}$ (top) and of
         $\alpha_{32}$ (bottom) for $^{96}$Zr (tetrahedral magic gap $N=56$).
         The results with the three Gogny interactions, D1S, D1N, and D1M are
         compared. The model space is defined by $N_{\rm osc}^{\rm max}=$12.
}
                                                                 \label{fig.09}
\end{center}
\end{figure}

In this section we discuss the results for the doubly-magic tetrahedral-symmetry nucleus $^{96}$Zr whose single-particle energy-gap properties and related diagrams have been discussed e.g.~in refs.~\cite{Dudetal14} and \cite{Schun04}.  The potential energy curves as functions of $\alpha_{20}$ for $^{96}$Zr calculated using the HFB and employing the three Gogny-type interactions are shown in fig.\,\ref{fig.09}, analogous to fig.\,\ref{fig.02} for $^{80}$Zr (in obtaining these results the angular-momentum projection has {\em not} been employed). It can be seen that the tetrahedral configuration has the lowest energy except for the D1N parametrisation, in which case the oblate-deformed minimum is slightly lower in energy. It is worth noticing that the potential energy curves with the three forms of parameterisation are considerably different in this nucleus. The single-particle energy gap at $N=56$ is not as large as the one at $N=40$ and the energy gain of the tetrahedral deformation is rather small as it is shown in the lower panel in fig.\,\ref{fig.09}.

However, the situation changes dramatically when the angular-momentum projection is included. The energy gain with the tetrahedral deformation becomes as large as 10 MeV; the projected energy curve of the $I^\pi =0^+$ state as a function of $\alpha_{32}$ is very similar to that in the case of $^{80}$Zr (not shown). The slightly smaller single-particle gap at $N=56$ implies the fact that the neutron pairing correlations survive at the tetrahedral minimum which leads to the
extra correlation energy by the number projection for neutrons, which is about $-1$ MeV. The numbers of mesh points employed for the Gaussian quadratures when constructing the projectors are
$
 (N_\alpha=N_\gamma,N_\beta)=(36,36)
$
for the three Euler angles, and $N_\phi=9$ for the gauge angle. The projection calculation from the superconducting mean-field states is much more time-consuming numerically, and this is also why a smaller model space with
$
 N_{\rm osc}^{\rm max}=8
$
has been used for this calculation.

\begin{figure}[!ht]
\begin{center}
\includegraphics[width=70mm]{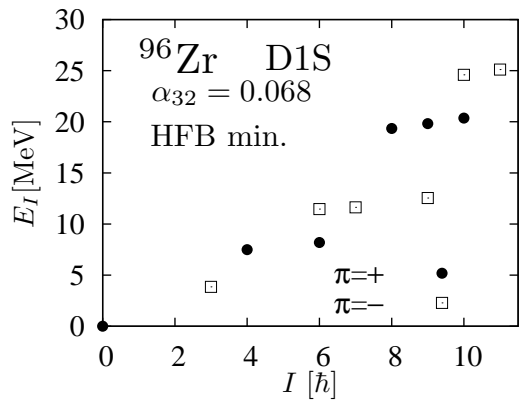}
\caption{
         The lowest energy levels calculated by projection from the HFB minimum state at
         $\alpha_{32}=0.068$ in $^{96}$Zr. The Gogny D1S parameterisation is
         used with the model space defined by $N_{\rm osc}^{\rm max}=8$.
         For comments about the characteristic features of these predictions see the text.
}
                                                                 \label{fig.10}
\end{center}
\end{figure}

The calculated lowest energy levels for $^{96}$Zr
are shown in fig.\,\ref{fig.10}, which are obtained by employing
full-projection techniques from the HFB minimum.
The HFB energy curve in fig.\,\ref{fig.09} is very shallow and the minimum deformation at $\alpha_{32}\approx 0.07$ might be considered small. However, the fact that the energy landscape is flat implies the existence of the low energy large amplitude motion with a possibly strong, in the present context the {\em dynamical} effect of the nuclear shape with the tetrahedral symmetry.

It is of interest to examine the structure of the energy spectrum with such a relatively small tetrahedral deformation. Let us notice that the lowest energy sequence is still given by the spin-parity combinations of eq.\,(\ref{eqn.05}), however this time the approximate degeneracies as the ones present in the case of $^{80}$Zr nucleus discussed earlier do not appear anymore. Instead, as it can be seen from the right-hand side of the figure, the $I^\pi = 0^+$, $3^-$, ($4^+$, $6^+$), ($6^-$, $7^-$, $9^-$), $\cdots$, states or groups of states are grouped resembling the structure of the zero, one, two, three, $\cdots$, -phonon multiplets, based on the elementary mode of the $3^-$ phonon. Although the anharmonicity is non-negligible, this kind of the phonon-like grouping pattern is characteristic of the near spherical tetrahedral spectra, see Ref.~\cite{Tagtetra13}.

The nucleus $^{96}$Zr is stable and many of its transitions have been known for many years, see e.g.~ref.\,\cite{TabIso96}. It is believed to be spherical, and its decay pattern contains a few sequences resembling the $I=(j^2)_I$-configurations as e.g.~$(j=\frac{7}{2})^2 \to 0^+, 2^+, 4^+, 6^+$. Indeed sequences of this type are rather abundantly seen in many spherical nuclei. However, there is no way of saying something about the shape of nuclei which involve the coupling of the $(A-2)$-nucleon core -- be it either spherical or, alternatively, (e.g.~tetrahedral) deformed, with the core-spin $I_{\rm core}=0$ -- and the above mentioned particle-hole type $(j^2)_{I}$-configurations. Any wave function $\Psi_{I_{\rm core}=0}$ will appear as spherical in the corresponding Clebsch-Gordan coupling independently of whether its underlying intrinsic moments are compatible with the spherical- or non-spherical (e.g.~tetrahedral) shape. In other words: there is no way of concluding about the underlying deformation or sphericity of the $(j^2)_{I}$-structures other then through a dedicated measurement of the related multipole moments in the sequence; this information being usually very scarce in the literature, some authors simply {\em assume} the sphericity rather then demonstrating it via experiment.

Trying to impose any non-trivial constraining conditions which would aim at pinning down the presence of the non-zero charge multipole moments would inevitably need to address the octupole $B(E3)$-transitions since the ideal tetrahedral-symmetric nucleus carries neither collective $E2$- nor $E1$-moments and for this very reason it will appear as spherical in any analysis which takes into considerations only the $\lambda=1$ and 2 multipolarity. But the $E3$-properties of the lowest $3^-$ state in $^{96}$Zr have been measured and its $B(E3)$ reduced transition probabilities of $(57 \pm 4)$ W.\,u.~are among the largest ever measured and stronger than those in e.g.~$^{208}$Pb. Moreover, the $N=56$ isotones have comparably strong reduced $B(E3)$ transition probabilities,
see table 1 and the discussion in ref.\,\cite{Dudetal14}, where the references to the experimental information can also be found.

The observed lowest $3^-$ state has an excitation energy 1.897 MeV, which is much lower than the value calculated within the present Gogny formalism, 3.85 MeV, using the projection from the state with $\alpha_{32}=0.068$. This suggests that the {\em calculated here} $3^-$ state is not the collective vibrational excitation in $^{96}$Zr -- but to draw conclusions about the vibrational nature of the discussed state we would need to extend the formalism e.g.~to include the particle-vibration coupling, what goes beyond the scope of this article.


\subsection{The case of the $^{110}_{\,\,\,40}$Zr$_{\,70}$ nucleus}
\label{Sect.03.04}

In this section we present the mechanism of competing symmetries, focussing on the tetrahedral symmetry for the case of the neutron-rich, unstable, tetrahedral doubly-magic nucleus $^{110}$Zr. Indeed, according to the previously published results, with the neutron number $N=70$, one obtains a big single particle gap at a significant $\alpha_{32}$-deformation, what allows for its qualification as the next tetrahedral-magic number, cf.~ref.\,\cite{Dudek02}.

\begin{figure}[!ht]
\begin{center}
\includegraphics[width=80mm]{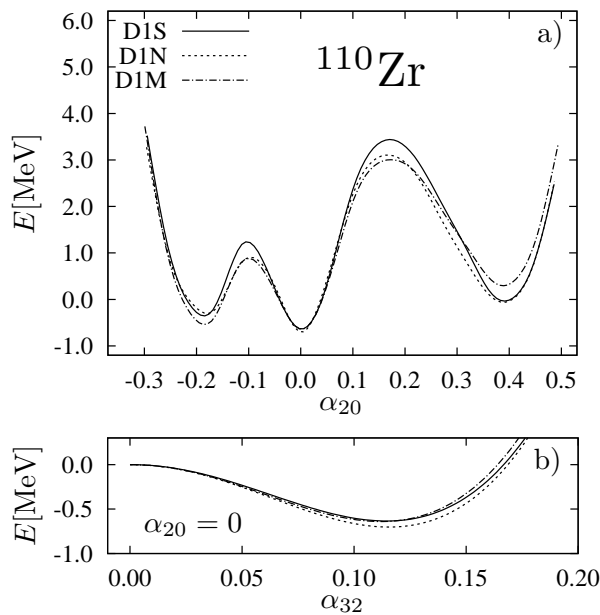}
\caption{
         The potential energy curves obtained with the constrained HFB
         calculation with projection on the good particle-numbers and parity,
         as functions of the deformation $\alpha_{20}$
         (top) and $\alpha_{32}$ (bottom) for $^{110}$Zr ($N=70$). The results
         correspond as before to the three choices of parameterisation of the
         Gogny interaction, D1S, D1N, and D1M. The model space is defined by
         $N_{\rm osc}^{\rm max}=$12.
}
                                                                 \label{fig.11}
\end{center}
\end{figure}

The low energy part of the excitation scheme of this nucleus has been studied in ref.\,\cite{Tagtetra13} employing the Woods-Saxon potential together with the simplified, separable two-body schematic interaction like the one used for the case of $^{80}$Zr (fig.\,\ref{fig.07}). Similarly as for the other isotopes discussed in this article, we compare the potential energy curves as functions of $\alpha_{20}$ calculated using D1S, D1N and D1M variants of the parametrisation of the Gogny interaction, cf.~fig.\,\ref{fig.11}.
All the curves are rather similar showing three minima along the $\alpha_{20}$-projection axis with the oblate ($\alpha_{20}\approx -0.2$), spherical (the lowest) and the prolate, nearly super-deformed quadrupole minimum at $\alpha_{20}\approx 0.4$. However, the spherical minimum in this representation turns out to be unstable with respect to the tetrahedral deformation, with the energy gain of about 0.7 MeV at the static tetrahedral deformation $\alpha_{32} \approx 0.12\,$. In contrast to the nucleus $^{80}$Zr, where the tetrahedral minimum is significantly lower in energy as compared to the other minima, in the present case all the three minima have rather similar energies, though with the significant separating potential barriers.

The effect of full projection for the $I^\pi =0^+$ state is also large for this nucleus and the energy gain relative to the spherical shape is about 10 MeV again (not shown), which is very similar to analogous results for $^{80}$Zr and $^{96}$Zr. The numbers of the mesh points employed for the Gaussian quadratures of the projectors are same as in the case of $^{96}$Zr. The pairing is non-vanishing for neutrons and, as before, we have used the smaller model space
($N_{\rm osc}^{\rm max}=$8).
\begin{figure}[!ht]
\begin{center}
\includegraphics[width=70mm]{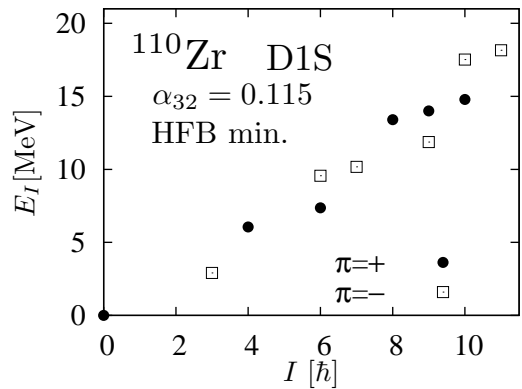}
\caption{
         The lowest energy levels calculated
         by projection from the HFB minimum state at
         $\alpha_{32}=0.115$ in $^{110}$Zr. The Gogny D1S
         parametrisation is used with the model space defined by
         $N_{\rm osc}^{\rm max}=8$. Let us notice that the pattern
         of the spin-parity combinations of this lowest-lying sequence resembles
         again the sequence of eq.\,(\ref{eqn.05}), see discussion following fig.\,\ref{fig.05}.
}
                                                                 \label{fig.12}
\end{center}
\end{figure}

Finally, the low-lying excitations energies for increasing spin calculated using the full-projection based on the HFB minimum are displayed in fig.\,\ref{fig.12} for $^{110}$Zr. In this nucleus the value of the tetrahedral static-equilibrium deformation for the HFB minimum, $\alpha_{32}=0.115$, is similar to that of $^{80}$Zr. Consequently, the two energy schemes look rather similar and because of the smallness of the tetrahedral deformation in both cases the energy vs.~spin dependence is approximately linear. However, the $^{110}$Zr nucleus is considerably heavier than the other Zirconium isotopes discussed in this article  and the excitation energy of the first $3^-$ state is less than 3 MeV; it is the smallest in sequence of the calculated isotopes. Compared to the calculations with the schematic Woods-Saxon plus multipole-multipole interactions of ref.\,\cite{Tagtetra13} the results with the Gogny D1S interaction used here are quantitatively similar, indicating that the collective excitations can be well described by the simple schematic Hamiltonians as it is shown in the case of $^{80}$Zr, see figs.\,\ref{fig.05} and~\ref{fig.07} (except for the vibrational states whose inclusion in the analysis would require a dedicated extension of the present formalism).


\section{Summary}
\label{Sect.05}

In the present article we have investigated the low-energy excitation patterns for doubly-magic tetrahedral-symmetry nuclei by employing the simultaneous particle-number, parity and angular-momentum projection method with the Gogny interaction. The pure tetrahedral-symmetry shape is expected to appear as the ground state for all the tetrahedral double magic Zr isotopes studied here: $^{80}$Zr$_{40}$, $^{96}$Zr$_{56}$, and $^{110}$Zr$_{70}$.
More precisely, calculations of the nuclear potential energy with the Gogny-HFB method show that only for $^{96}$Zr and with the Gogny D1N interaction the lowest energy state is not of tetrahedral symmetry, whereas all the others are.

The energy gain in terms of increasing tetrahedral deformation is rather weak with the Gogny-HFB method, depending slightly on the actual parameterisation used, and amounts to only a few hundred keV for $^{96}$Zr and less than one MeV for $^{80}$Zr and $^{110}$Zr. However, the effects of possibly small static tetrahedral deformations must not be ignored. They are usually accompanied with the flat energy landscapes in the $\alpha_{32}$-direction implying large amplitude oscillations and generating possibly large dynamical presence of the tetrahedral symmetry in the considered nuclei.
Among the three Zr isotopes, $^{80}$Zr has the most stable tetrahedral minimum,
i.e., the energy difference between the tetrahedral and other minima is the
largest and exceeds 3 MeV.

When the effect of the angular momentum projection is included the situation dramatically changes. The correlation energy of the projection is very large for
the octupole deformation including the tetrahedral one ($\alpha_{32}$),
typically about 10~MeV for Zr isotopes, which is much larger than that for the quadrupole deformation. Therefore the energy landscapes of the projected
$I^\pi =0^+$ states change very importantly.
In particular, it should be expected that the calculated energy minima
associated with the tetrahedral shapes will correspond to larger tetrahedral deformation when the angular momentum projection is applied.
The spectrum for the collective excitations based on the tetrahedral shape
is also calculated by the projection method.
The calculated HFB minimum corresponds to rather small tetrahedral deformation, $\alpha_{32}\approx 0.07-0.12$, and the energy vs.~spin dependence is nearly linear - what constitutes a possible element of the experimental testing of the discussed properties.

In the present work we have focussed on the microscopic calculations of the low-energy spectra in nuclei with the pure tetrahedral deformation (pure tetrahedral symmetry). However, more generally, some other deformations leading to lowering of the total nuclear energy would be superimposed, e.g., triaxial quadrupole deformation and/or the pear shape ($\alpha_{30}$) deformation. In order to study such a more general case, one may like to combine
the full-projection and configuration mixing.
Such an extension is under progress.

Last but not least, we could fully confirm the very specific result that the lowest energy sequence of states either forming the yrast line or lying very close to the yrast states has the spin-parity combinations characteristic of the $A_1$ irreducible representation of the tetrahedral point-group, $T_d$. All other, `next-lowest lying states', are predicted to form the sequences approximately parallel but appearing significantly (several MeV) higher in the energy scale. This grouping of states seems to indicate again the usefulness of the information coming from the theory of representations of the symmetry groups of the mean-field Hamiltonians. However, these predictions apply only to the states which can be interpreted as rotational ones, associated with the tetrahedral-symmetry minima; such states are expected to compete with e.g.~particle-hole non-collective excitations as well as vibrational collective states.


\section*{Acknowledgements}

This work is supported in part by Grant-in-Aid for Scientific Research
No.~25$\cdot$949 and No.~22540285
from Japan Society for the Promotion of Science,
and by the Polish-French COPIN collaboration under project number 04-113.


\end{document}